# Spin mediated enhanced negative magnetoresistance in $Ni_{80}Fe_{20}$ and p-silicon bilayer

Paul C Lou[1] and Sandeep Kumar[1,2,*]

[1] Department of Mechanical Engineering, University of California, Riverside, CA

[2] Material Science and Engineering Program, University of California, Riverside, CA

Abstract

In this work, we present an experimental study of spin mediated enhanced negative magnetoresistance in $Ni_{80}Fe_{20}$ (50 nm)/p-Si (350 nm) bilayer. The resistance measurement shows a reduction of ~2.5% for the bilayer specimen as compared to 1.3% for $Ni_{80}Fe_{20}$ (50 nm) on oxide specimen for an out-of-plane applied magnetic field of 3T. In the $Ni_{80}Fe_{20}$-only film, the negative magnetoresistance behavior is attributed to anisotropic magnetoresistance. We propose that spin polarization due to spin-Hall effect is the underlying cause of the enhanced negative magnetoresistance observed in the bilayer. Silicon has weak spin orbit coupling so spin Hall magnetoresistance measurement is not feasible. We use $V_{2\omega}$ and $V_{3\omega}$ measurement as a function of magnetic field and angular rotation of magnetic field in direction normal to electric current to elucidate the spin-Hall effect. The angular rotation of magnetic field shows a sinusoidal behavior for both $V_{2\omega}$ and $V_{3\omega}$, which is attributed to the spin phonon interactions resulting from the spin-Hall effect mediated spin polarization. We propose that the spin polarization leads to a decrease in hole-phonon scattering resulting in enhanced negative magnetoresistance.

Silicon is an appealing material for spintronics because it has a relatively weak spin-orbit coupling, crystal inversion symmetry and zero nuclear spin; hence, it has no spin relaxation due to scattering, Dyakonov-Perel spin relaxation mechanism and hyperfine interactions[1]. Spin injection in silicon can be achieved by tunneling from a ferromagnetic electrode across a thin insulator[2-7]. Spin injection in silicon has also been demonstrated using spin Seebeck tunneling[8]. Spin-Seebeck effect[9] studies on permalloy/GaMnAs[10, 11] clearly suggest phonon-driven spin redistribution. The critical question that arises from these studies is the role of spin polarization on transport behavior. The spin diffusion length in silicon has been measured between 250 nm and 6 μm. At room temperature, electron-phonon scattering (Elliot-Yafet mechanism) is believed to be the primary spin relaxation mechanism. We hypothesize that spin injection may change the charge carrier scattering behavior, especially with phonons, leading to changes in charge transport behavior. This prompted us to study the charge transport behavior in silicon due to spin polarization.

To test our hypothesis, we developed an experimental setup to observe the changes in electrical transport behavior. We fabricated a MEMS-based setup having a $Ni_{80}Fe_{20}$/p-type silicon bilayer specimen with four-point probe geometry for resistance measurement. To fabricate the MEMS device, we took a silicon-on-insulator (SOI) wafer with a highly doped (B) p-type 2 μm thick device layer having resistivity of 0.001-0.002 Ω cm. At room temperature, the spin diffusion length for p-type Si is reported to be 310 nm [4]. We chemically etched the device layer of the SOI wafer to achieve the thickness closer to ~350 nm, near the spin diffusion length, by successively oxidizing and etching the wet thermal oxide using hydrofluoric (HF) acid. The surface oxide from the silicon was removed by Ar milling for 15 minutes followed by deposition of a layer of 50 nm $Ni_{80}Fe_{20}$ using RF sputtering. Using UV photolithography, specimen and contact pads were patterned on the wafer. The $Ni_{80}Fe_{20}$ was wet-etched with 1:8 nitric acid-water solution, and Si was etched with inductively coupled plasma using $SF_6$ and $CF_4$. A scanning electron micrograph (SEM) of the fabricated device is shown in Figure 1. We also fabricated devices having $Ni_{80}Fe_{20}$ thin film on silicon oxide as control devices. The devices were wire-bonded and placed inside the Quantum Design Physical Property Measurement System for the charge transport study on this bilayer. Electrical resistance measurement was carried out using delta mode with a Keithley 6221 current source and 2182A nanovoltmeter. The delta mode utilizes the current reversal technique to cancel out any constant thermoelectric offsets, so the results reflect the true value of the voltage being measured. The magnitude of the current used for measurement is 10 µA. We measured the resistance of the bilayer as ~114.6 Ω and ~153.7 Ω for $Ni_{80}Fe_{20}$ thin films. We approximate the resistivity as 40 µΩ cm, which is slightly higher than the value reported for 50 nm films[12]. Assuming parallel resistors for bilayer geometry, the resistance of the p-Si layer is calculated to be ~450 Ω.

(Figure 1)

We did the preliminary (coarse) measurement of the resistance (MR) as a function of the magnetic field (Supplementary Figure S1). From this data, we observe a reduction in electrical resistance by 2.5% in the case of out-of-plane magnetic field of 3 T, and a reduction of 1.15% for the transverse in-plane magnetic field of 3 T. In order to quantify the contribution of p-Si to the observed behavior, we tested the fabricated control $Ni_{80}Fe_{20}$ thin film device, in which the $Ni_{80}Fe_{20}$ layer is in contact with $SiO_2$ rather than Si. We observed a negative MR of 1.3% in 50 nm $Ni_{80}Fe_{20}$ thin film for the transverse out-of-plane magnetic field and a 0.45% change for the transverse in-plane magnetic field. The negative MR in $Ni_{80}Fe_{20}$ thin films has been observed for thicknesses below 100 nm; this is attributed to magnetic anisotropy or anisotropic magnetoresistance (AMR)[12, 13]. The slope of MR after ~1.25T (out of plane) in case of bilayer is same as $Ni_{80}Fe_{20}$ on oxide. From these observations on $Ni_{80}Fe_{20}$ thin films, we can deduce that the negative MR observed in our bilayer specimen has contributions from the $Ni_{80}Fe_{20}$ layer as well as the Si layer. In addition, the preliminary magnetoresistance measurement does not show the magnetic field dependent behavior at low magnetic field, which is essential to discover the underlying mechanism of observed behavior.

(Figure 2)

We propose that the observed enhanced negative MR in the bilayer is due to spin polarization of holes in the p-Si layer. Dash et al.[4, 6] measured the spin diffusion lengths as 310 nm in p-type Si and 230 nm in n-type Si. Experimental evidence suggest that holes tend to have longer spin diffusion lengths[14]. The thickness of the Si layer in this study is approximately 350 nm. We propose that the spin diffusion length in the present experiment is higher than the Si layer thickness (350 nm), inducing complete spin polarization in the Si layer. To test this assertion, we measured the $R_{1\omega}$, $V_{2\omega}$ and $V_{3\omega}$ as a function of applied transverse and out of plane magnetic field from 1500 Oe to -1500 Oe as shown in Figure 2. In addition, course magnetic field dependent measurement may induce significant error in the in-plane magnetoresistance due to remnant magnetic field, which necessitates the experimental measurement of the low field magnetoresistance behavior. The low field negative magnetoresistance is found to be 0.88% for the $Ni_{80}Fe_{20}$ control specimen and 1.78% for $Ni_{80}Fe_{20}$

/p-Si bilayer specimen. Since the zero field resistance is different for both the specimen, we analyzed the change resistance (ΔR) as shown in Figure 2 (a) and (c). We observe an enahcement in negative magnetoresistance for bilayer specimen. Further, for the $Ni_{80}Fe_{20}$ control specimen, the resistance peaks at zero field, whereas for the bilayer specimen, the peak appears at different points in the increasing and decreasing magnetic fields. This pseudo spin valve supports our hypothesis that spin polarization in the p-Si layer is the underlying mechanism for the observed behavior. For the out of plane magnetic field, we observe an exchange bias and switching behavior as show in Figure 2. This behavior might arise from the interfacial antiferromagnetic layer.

(Figure 3)

The observed spin polarization may occur due to thermally driven spin-Seebeck tunneling from the $Ni_{80}Fe_{20}$ to the Si layer The thermally-driven spin injection method is believed to have longer spin diffusion length relative to conventional spin injection, since thermal redistribution of spin does not depend on transport of charge[8]. As the electric current is applied to the bilayer, a temperature gradient sets in due to differential Joule heating of both the layers (the electrical conductivity of the $Ni_{80}Fe_{20}$ thin film layer is significantly larger than that of the p-Si layer). Thermal conductivity differences should also contribute to the thermal gradient; $Ni_{80}Fe_{20}$ has a thermal conductivity of 23 W/mK (approximately for ~50 nm thick[15]), which is significantly lower than that of Si ( ~80 W/mK[16]). This temperature gradient (see Figure 1 b) may cause the spin-Seebeck tunneling and spin polarization of holes in the Si layer. The spin-Seebeck tunneling occurs due to the difference between the electron temperature in

$Ni_{80}Fe_{20}$ and the phonon temperature in Si[17]. To verify the spin Seebeck tunneling, we measured the $V_{2\omega}$ as a function of magnetic field for bilayer and control specimen as shown in figure 3. We observe a large longitudinal $V_{2\omega}$ in the $Ni_{80}Fe_{20}$/p-Si bilayer specimen indicating existence of spin current. In addition, the $V_{2\omega}$ does not change sign when the sign of the magnetic field is reversed. This clearly demonstrates that the spin Seebeck effect is negligible in the specimen. We can rule out the spin-Seebeck tunneling as a mechanism responsible for spin polarization. We hypothesize that the reason of large $V_{2\omega}$ is spin current from the p-Si layer to $Ni_{80}Fe_{20}$ layer.

We propose that spin-Hall effect (SHE) in p-Si could be the other mechanism responsible for spin polarization. Recent observation of the inverse spin-Hall effect (ISHE) in p-Si[18] help support our observation. To ascertain the SHE, we measured the $R_{1\omega}$, $V_{2\omega}$ and $V_{3\omega}$ as the applied magnetic field is rotated in yz-plane as shown in Figure 4 (a)-(c) starting with magnetic field pointing in the z-direction. The measurement is carried out for constant magnetic fields of 1 T and 2 T. From Figure 4, the change in first harmonic resistance clearly shows the distorted $\sin^2\phi_{yz}$ dependence similar to anisotropic magnetoresistance (AMR)[19] behavior instead of spin Hall magnetoresistance (SMR). This AMR is attributed to the dimensionality effect in $Ni_{80}Fe_{20}$ thin films, causing out-of-plane AMR to be larger than the in-plane AMR. In addition, the magnetic moment shows a fast switching and then flat top. This behavior is attributed to the antiferromagnetic interfacial layer as stated earlier. Negative transverse MR in n-doped Si has been observed at low temperatures[20-22]. This phenomenon in n-Si MOS inversion layers is proposed to be caused by inelastic scattering due to electron-electron interactions in a weak localization regime. Schoonus et al. observed large positive MR in B-doped Si-$SiO_2$-Al structures with doping levels similar to those in the present work[23].

However, we observe negative MR in p-Si having a ferromagnetic spin source ($Ni_{80}Fe_{20}$) instead. They proposed that applying a magnetic field leads to an increase of the acceptor level compared with the valence band. We propose that the mechanism responsible for the observed negative MR is similar. The applied magnetic field causes an increase in acceptor states in p-Si (B-doped). The applied magnetic field along with the large doping concentration may lift the degeneracy of valence band maxima[24, 25]. The mechanical stress due to impurity (dopant) agglomeration also lifts the degeneracy of valence band maxima between a heavy hole and a light hole[26]. However, positive MR is observed in p-Si without a $Ni_{80}Fe_{20}$ spin source at room temperature[26]. This means that broken degeneracy of the valence band may be the necessary condition – but not a sufficient condition – for the observed negative MR.

We propose that spin polarization of p-Si due to SHE is essential for the observed behavior. In order to support our argument, we analyzed the angular variation of $V_{2\omega}$ and $V_{3\omega}$ as a function of magnetic field as shown in Figure 4 (b)-(c). We observe a $\sin^2\phi_{yz}$ response similar to SMR. In case of Si, the small spin orbit coupling lead to negligible spin Hall angle as states earlier. Hence, SMR like behavior is not observed in the first harmonic resistance. The $V_{2\omega}$ and $V_{3\omega}$ are related to the temperature fluctuations and thermal transport respectively. The spin phonon relaxation is the primary relaxation mechanism and phonons are the primary heat carrier in Si. We propose that spin phonon interactions leads to observed angular dependence in case of $V_{2\omega}$ and $V_{3\omega}$. The thermal conductivity is calculated using the following equation: $\kappa \approx \frac{4I^3 R_0 R' L}{\pi^4 V_{3\omega} S}$, where $\kappa$ is the thermal conductivity, $I$ is rms AC bias, $R_0$ is the initial electrical resistance of the specimen, $R'$ is the resistance derivative of temperature $R' = \left(\frac{dR}{dT}\right)_{T_0}$, $L$ is the length between the voltage contacts, $S$ is the cross-section area of the specimen, and $V_{3\omega}$ is the third harmonic voltage component. In this study, the specimen is not freestanding which may cause the conduction heat loss to the substrate. Due to the substrate heat loss, we may not be able to calculate the thermal conductivity but the $V_{3\omega}$ can still describe the thermal transport behavior of the specimen at low frequency measurement. The magnetic field dependent $V_{3\omega}$ clearly demonstrate the change in thermal transport as shown in Figure 4 (d). In order to ascertain the origin of this behavior, we measured the $V_{3\omega}$ for $Ni_{80}Fe_{20}$ on oxide as a function of magnetic field (Supplementary Figure S2), which does not show magnetic field dependence. We can clearly state that the origin of change angular variation and magnetic field dependent variation in $V_{3\omega}$ is due to spin polarization in p-Si. The spin phonon interactions are related to the spin polarization, which arises because of SHE. We propose that the observed enhanced negative magnetoresistance is attributed to the spin-phonon interactions due to SHE.

In conclusion, we presented experimental measurements of negative MR in a $Ni_{80}Fe_{20}$ (50 nm)/p-Si bilayer. The resistance measurement shows a reduction of ~2.5% for the bilayer specimen as compared to 1.3% for $Ni_{80}Fe_{20}$ (50 nm) on oxide specimen for an out-of-plane applied magnetic field of 3T. We propose that the SHE in $Ni_{80}Fe_{20}$/Si bilayer leads to spin polarization in the Si layer. The applied magnetic field lifts the degeneracy of valence band maxima. The spin dependent hole-phonon scattering behavior changes due to spin-phonon interactions. Thus, the spin-phonon interactions are proposed responsible for the observed behavior.

**References**

[1] Y. Song, H. Dery, Physical Review B, 86 (2012) 085201.
[2] A. Dankert, R.S. Dulal, S.P. Dash, Sci. Rep., 3 (2013).

[3] B.T. Jonker, G. Kioseoglou, A.T. Hanbicki, C.H. Li, P.E. Thompson, Nat Phys, 3 (2007) 542-546.

[4] R. Jansen, Nat Mater, 11 (2012) 400-408.

[5] S. Zhang, S.A. Dayeh, Y. Li, S.A. Crooker, D.L. Smith, S.T. Picraux, Nano Letters, 13 (2013) 430-435.

[6] S.P. Dash, S. Sharma, R.S. Patel, M.P. de Jong, R. Jansen, Nature, 462 (2009) 491-494.

[7] I. Appelbaum, B. Huang, D.J. Monsma, Nature, 447 (2007) 295-298.

[8] J.-C. Le Breton, S. Sharma, H. Saito, S. Yuasa, R. Jansen, Nature, 475 (2011) 82-85.

[9] K. Uchida, S. Takahashi, K. Harii, J. Ieda, W. Koshibae, K. Ando, S. Maekawa, E. Saitoh, Nature, 455 (2008) 778-781.

[10] C.M. Jaworski, J. Yang, S. Mack, D.D. Awschalom, J.P. Heremans, R.C. Myers, Nat Mater, 9 (2010) 898-903.

[11] C.M. Jaworski, J. Yang, S. Mack, D.D. Awschalom, R.C. Myers, J.P. Heremans, Physical Review Letters, 106 (2011) 186601.

[12] T.G.S.M. Rijks, S.K.J. Lenczowski, R. Coehoorn, W.J.M. de Jonge, Physical Review B, 56 (1997) 362-366.

[13] A.O. Adeyeye, G. Lauhoff, J.A.C. Bland, C. Daboo, D.G. Hasko, H. Ahmed, Applied Physics Letters, 70 (1997) 1046-1048.

[14] E. Shikoh, K. Ando, K. Kubo, E. Saitoh, T. Shinjo, M. Shiraishi, Physical Review Letters, 110 (2013) 127201.

[15] A.D. Avery, S.J. Mason, D. Bassett, D. Wesenberg, B.L. Zink, Physical Review B, 92 (2015) 214410.

[16] M. Asheghi, K. Kurabayashi, R. Kasnavi, K.E. Goodson, J. Appl. Phys., 91 (2002) 5079-

5088.

[17] M. Schreier, N. Roschewsky, E. Dobler, S. Meyer, H. Huebl, R. Gross, S.T.B. Goennenwein, Applied Physics Letters, 103 (2013) 242404.

[18] K. Ando, E. Saitoh, Nat Commun, 3 (2012) 629.

[19] T.G.S.M. Rijks, S.K.J. Lenczowski, R. Coehoorn, W.J.M. de Jonge, Physical Review B, 56 (1997) 362-366.

[20] I. Eisele, G. Dorda, Physical Review Letters, 32 (1974) 1360-1363.

[21] Y. Kawaguchi, S. Kawaji, Journal of the Physical Society of Japan, 48 (1980) 699-700.

[22] Y. Kawaguchi, S. Kawaji, Surface Science, 113 (1982) 505-509.

[23] J.J.H.M. Schoonus, F.L. Bloom, W. Wagemans, H.J.M. Swagten, B. Koopmans, Physical Review Letters, 100 (2008) 127202.

[24] P.Y. Yu, M. Cardona, Fundamentals of semiconductors, Springer, 2005.

[25] H. Tokumoto, T. Ishiguro, R. Inaba, K. Kajimura, K. Suzuki, N. Mikoshiba, Physical Review Letters, 32 (1974) 717-720.

[26] A. Onton, P. Fisher, A.K. Ramdas, Physical Review, 163 (1967) 686-703.

List of Figures:

Figure 1. (a) SEM micrograph of experimental MEMS device, (b) schematic showing temperature gradient and spin diffusion in bilayer specimen, MR measurement for 50 nm $Ni_{80}Fe_{20}$(on oxide) and 50 nm $Ni_{80}Fe_{20}$/~350 nm p-type silicon bilayer specimens (c) transverse magnetic field and (d) out-of-plane magnetic field.

Figure 2. Resistance as a function of magnetic field for 50 nm $Ni_{80}Fe_{20}$ thin film (a) y-dir, (b) z-dir, and 50 nm $Ni_{80}Fe_{20}$/~350 nm p-type silicon bilayer (c) y-dir and (d) z-dir

Figure 3. $V_{2\omega}$ as a function of magnetic field for 50 nm $Ni_{80}Fe_{20}$ thin film (a) y-dir, (b) z-dir, and 50 nm $Ni_{80}Fe_{20}$/~350 nm p-type silicon bilayer (c) y-dir and (d) z-dir

Figure 4. (a) $R_{1\omega}$, (b) $V_{2\omega}$ and (c) $V_{3\omega}$ as function of magnetic field rotation in *yz*-plane, and (d) the $V_{3\omega}$ as a function of applied magnetic field for transverse in-plane (y-dir) and out of plane (z-dir) magnetic field.

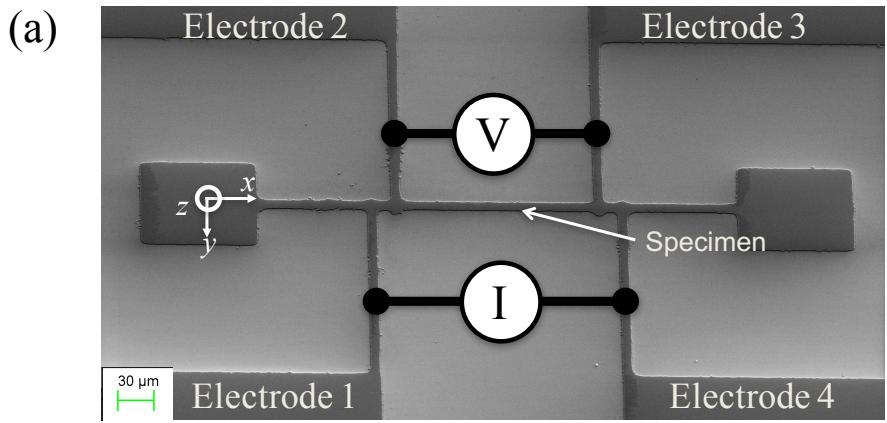

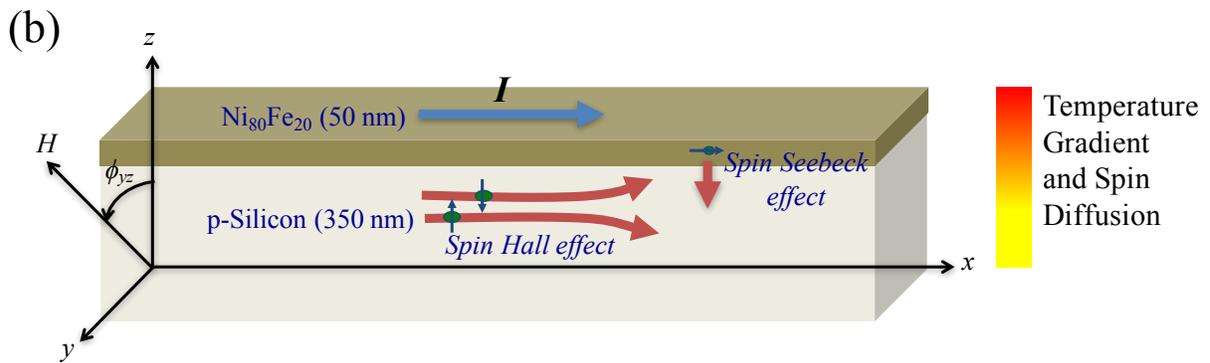

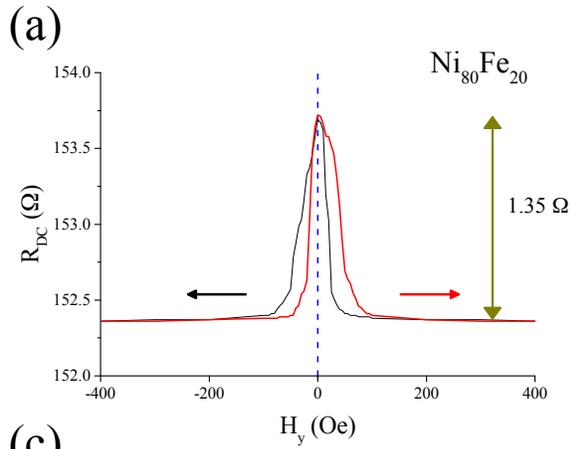
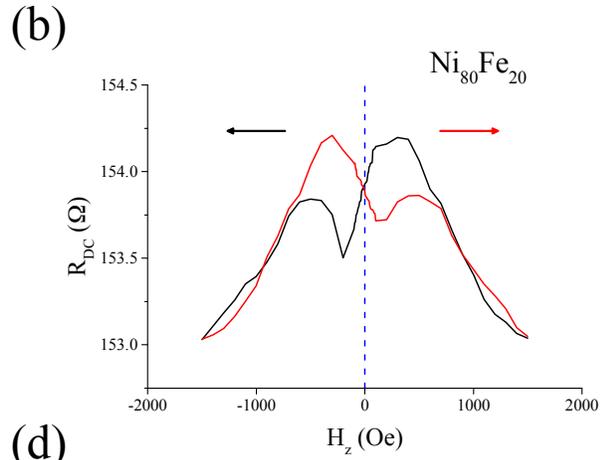
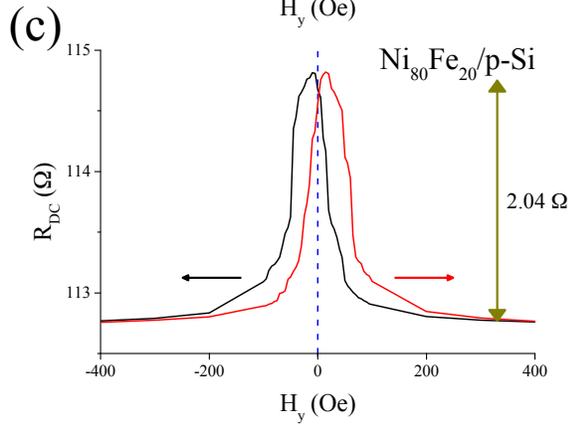
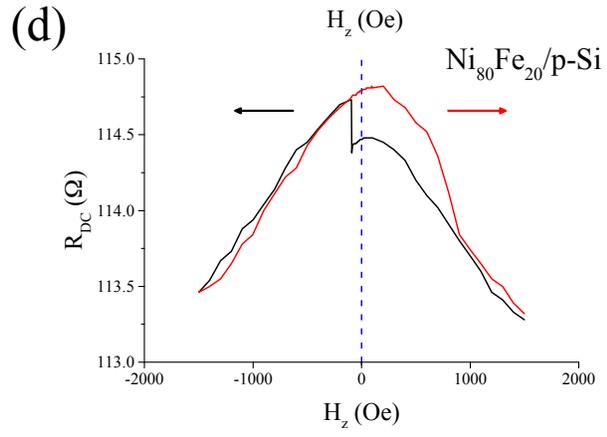

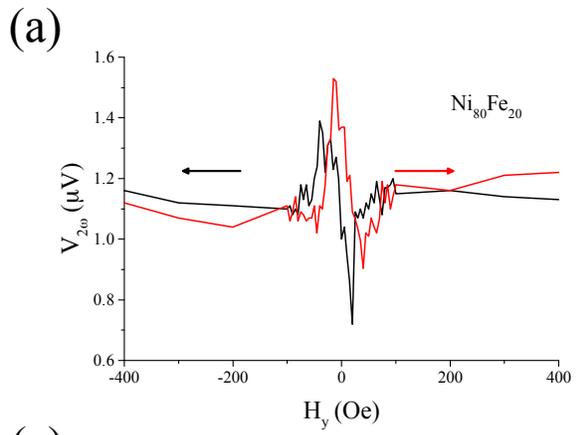
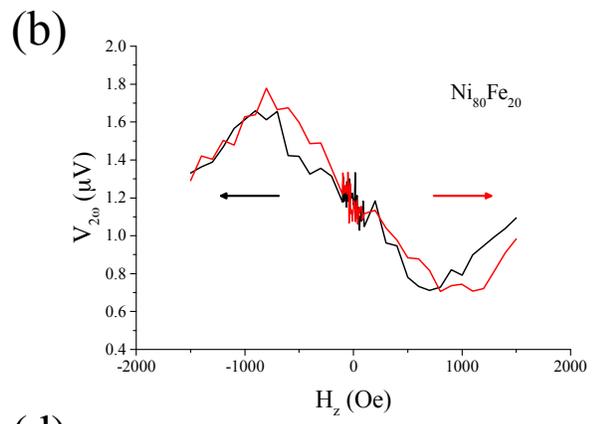
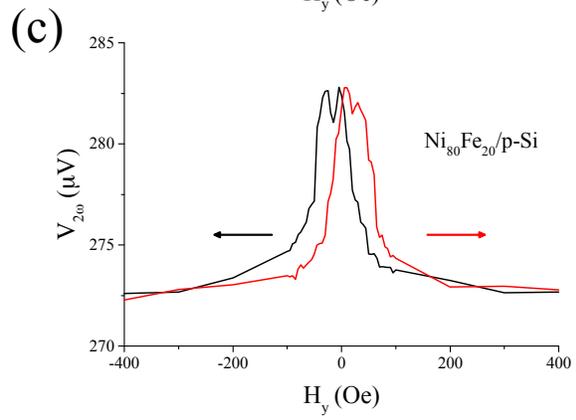
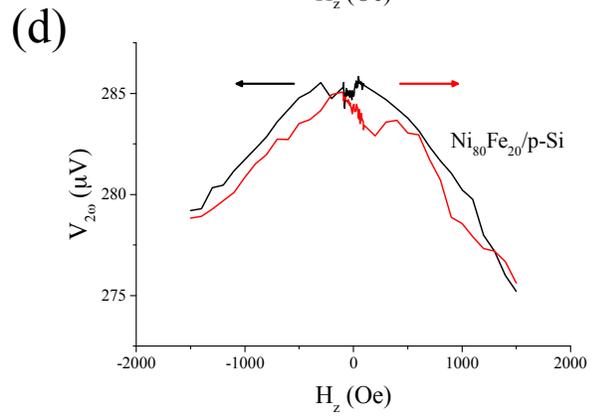

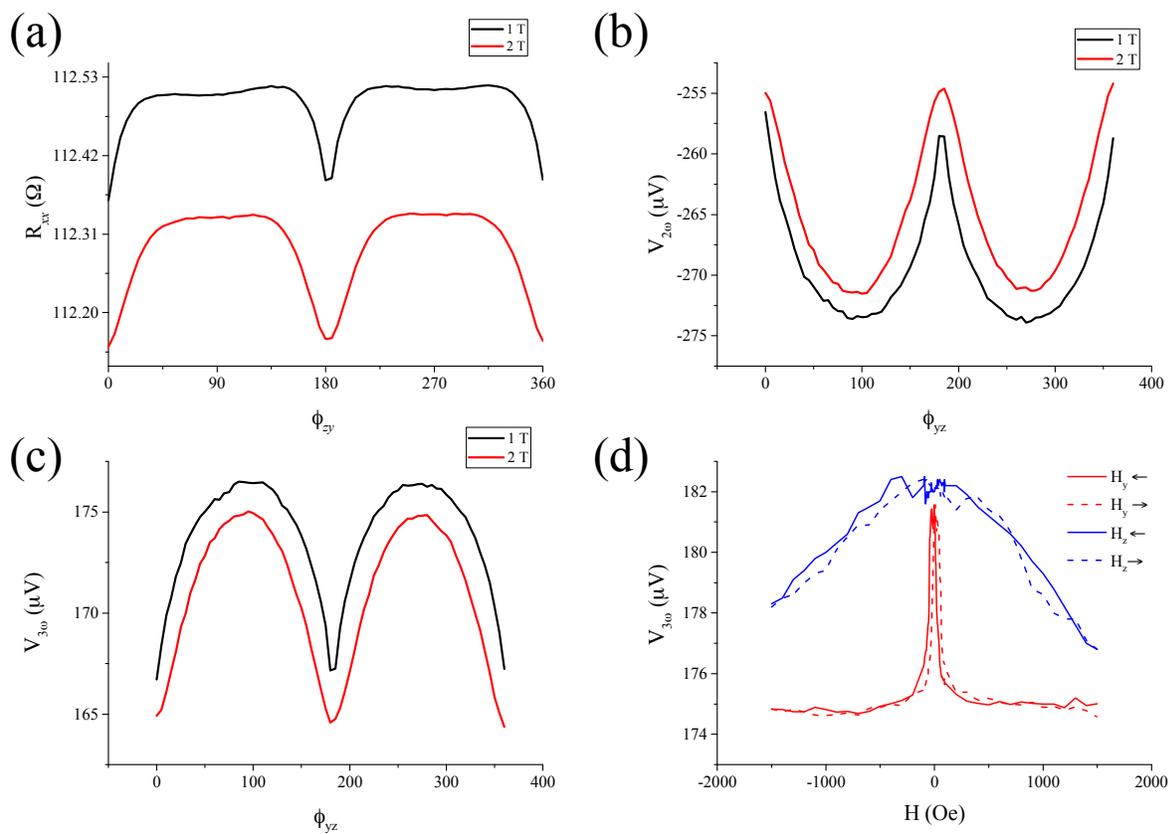

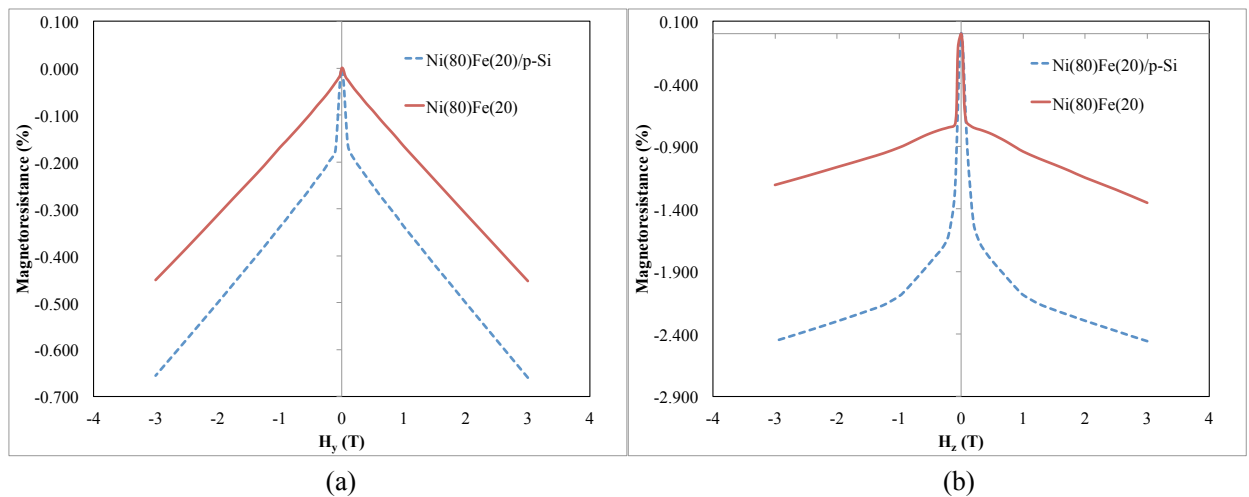

Supplementary Figures-

Figure S1. MR measurement for $Ni_{80}Fe_{20}$ and bilayer specimens (a) transverse magnetic field and (b) out-of-plane magnetic field.

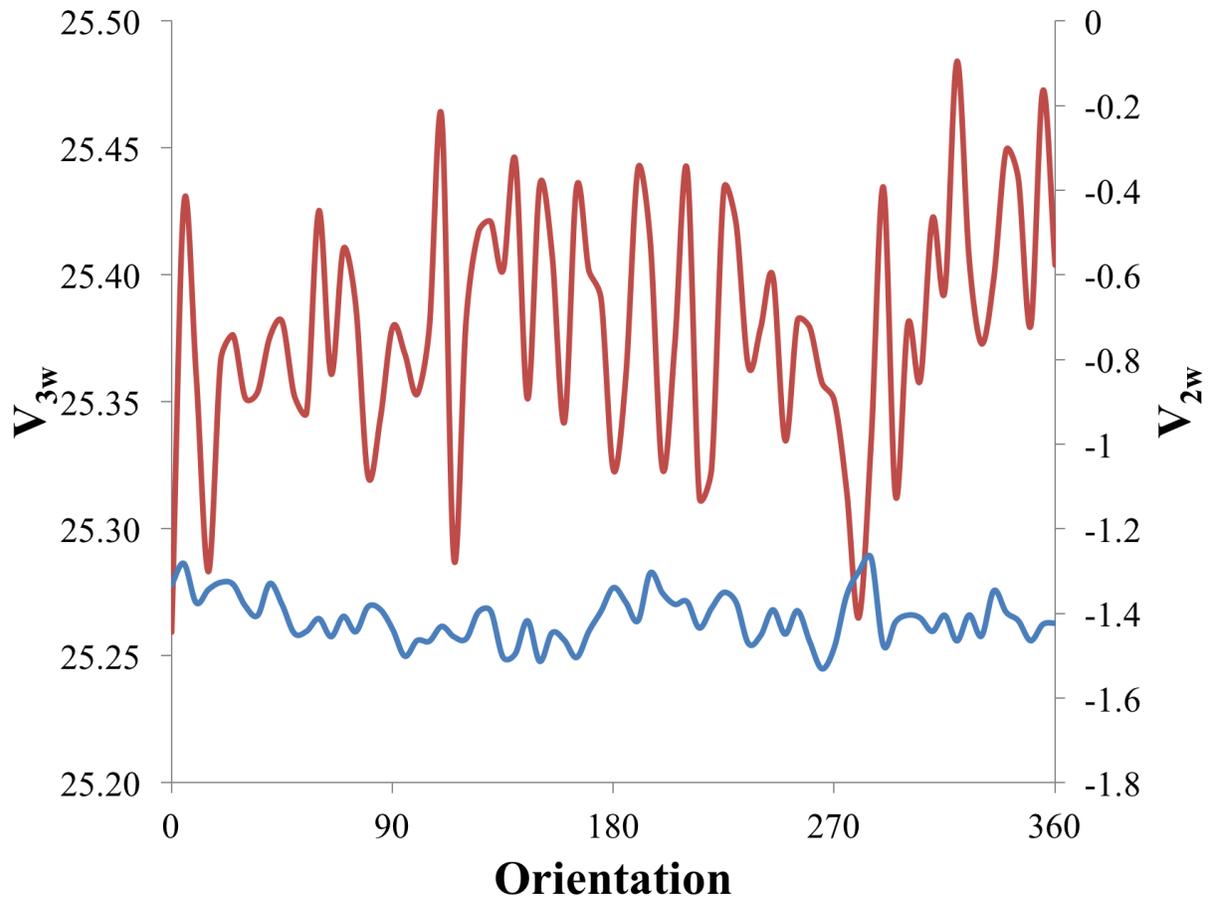

Figure S2. The $V_{2\omega}$ and $V_{3\omega}$ as a function of angular rotation of magnetic field in yz-plane for $Ni_{80}Fe_{20}$ specimen.